# Efficient Tunable THz wave Generation using Spectral Shaping in Electro-Optic Combs

Sanghoon Chin, and Ewelina Obrzud

*Abstract*— We propose a novel method for generating terahertz (THz) waves using cascaded electro-optic modulators (EOMs), providing numerical simulations and experimental validations to demonstrate its effectiveness. The key of this technique is to apply hybrid amplitude and phase modulation to the incident light, which significantly enhances the control capability over the output optical spectrum. This enhanced control turns to be crucial for the efficient generation of THz waves. By carefully adjusting the relative phase between microwave RF signals applied to the amplitude and phase modulators and by applying an appropriate DC bias to the amplitude modulator, we can considerably reduce the optical power of unwanted sidebands. This allows us to concentrate more energy on desired frequencies, facilitating the generation of high-purity THz waves through optical heterodyne mixing at the detector. To validate our concept, we have successfully generated a dual-frequency-like laser with as spectral separation of 0.897 THz. This result was accomplished by modulating the phase EOM at 34.5 GHz with a modulation depth of ~4.7, paving the way for the generation of more practical and efficient THz waves.

## I. INTRODUCTION

THE generation of microwave and terahertz (THz) waves has been extensively investigated and rapidly advanced due to recent developments in microwave photonics technology. These advancements enhance multiple complex functions in radio-frequency engineering and create novel opportunities through the wide bandwidth and low loss of modern photonics [1]. Photonic techniques for generating THz wave signals have garnered significant attention in the terahertz science community due to their unique properties. For instance, thanks to their short wavelengths, THz waves are less susceptible to free-space diffraction and interference with neighboring antennas. In addition, they offer high resilience to eavesdropping and greater link directionality in compact setups. Such advantages make the THz communication ideal for providing large bandwidth and high-capacity services over long distances, meeting the demands of future sixth-generation communication and beyond [2]. Another vital application of THz waves lies in their unmatched capability for non-destructive and non-invasive imaging. THz wave radiation is non-ionizing and can penetrate materials such as plastic, wood, paper, and textile. Consequently, THz imaging sensors can accurately diagnose and screen objects in non-metallic containers without causing damage, thanks to their low photon energy [3].

The most straightforward photonic scheme for generating microwave and THz band signals is based on directly modulated semiconductor lasers. Electrical signals applied to the laser can be easily converted into optical signals to produce multiple harmonics of the applied modulation frequency. However, the maximum achievable frequency of generated microwave signals is practically limited, typically to tens of GHz due to the modulation rate and bandwidth of the employed laser [4,5]. An alternative method leverages the optical heterodyne technique. The mutual interference between two separately operating lasers produces a heterodyne beat signal at the photodetector output. This way the beat frequency over the terahertz window can be readily obtained since it is essentially determined by the differential optical frequency between the two lasers [6,7]. However, this scheme is hindered by the instability of the frequency interval between the two lasers and random optical phase fluctuations, necessitating a complex regulation procedure for phase de-correlation compensation. To overcome this issue, the two-color laser concept has also been actively studied to generate THz wave signals [8-10].

Among existing photonic methods for microwave and THz waves generation, schemes utilizing optical frequency comb have been considered a breakthrough solution, demonstrating promising performance [11-15]. The combination of a single laser and multiple electro-optic modulators (EOMs) in series provides a straightforward and economical architecture to effectively produce a wideband frequency comb covering the THz range [16-22]. In turn, the direct optical heterodyne between optical comb lines through a photo-mixer in uni-traveling carrier photodiodes or high-speed p-i-n photodiodes enables the generation of highly stable THz carriers.

In this paper, we propose a simple and robust technique for generating very high-frequency carriers in the range of microwave and terahertz (THz) waves. This method utilizes spectral tailoring of the optical spectrum that is generated by sinusoidal hybrid phase and amplitude modulation. The optical phase of a seed laser operating in the C-band is intensively modulated using a phase EOM, resulting in a broadband optical frequency comb with line spacing equal to the modulation frequency applied to the modulator. The

This paragraph of the first footnote will contain the date on which you submitted your paper for review, which is populated by IEEE. This work was supported by the internal grant of Centre Suisse d'Electronique et de Microtechnique. *(Corresponding author: Sanghoon Chin).*

The authors are with the Business Unit of Instrumentations, Centre Suisse d'Electronique et de Microtechnique, Neuchâtel 2000, Switzerland (e-mail: sanghoon.chin@ csem.ch;ewelina.obrzud@csem.ch).

Color versions of one or more of the figures in this article are available online at http://ieeexplore.ieee.org

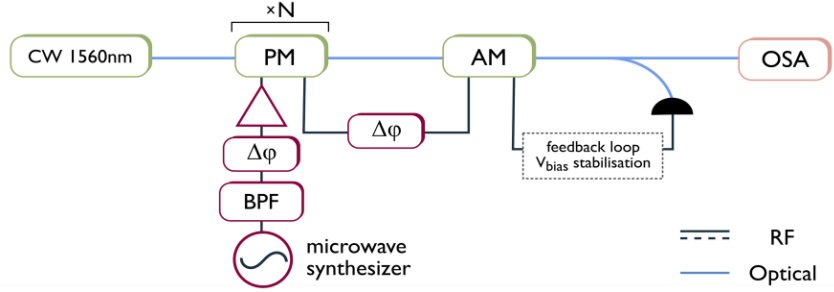

**Fig. 1** Simplified schematic diagram to generate wideband sidebands using sinusoidal hybrid modulation. CW: continuous wave. OFC: optical frequency comb

generated multicarrier is then fed into an amplitude-EOM to tailor the optical power of comb lines around the seed laser frequency. In most use cases, the role of amplitude-EOM is to generate a flat-top optical frequency comb. But, in our experiment, the parameters of amplitude modulator are optimized to suppress a large number of sidebands around the center frequency. This way, for the proof-of-concept, a dual-frequency-like light source with a spectral distance of 0.897 THz has been successfully generated by modulating the phase- and amplitude-EOM at 34.5 GHz. Notice that the relative frequency of the two tones is essentially a function of the RF power and/or frequency applied to the modulators, showing the tunability of the carrier frequency of the generated THz wave.

## II. PRINCIPLE OF SPECTRAL TAILORING IN ELECTRO-OPTIC FREQUENCY COMBS

Fig. 1 shows the simplified schematic diagram of conventional optical setup designed to generate wideband flat-top sidebands using sinusoidal hybrid modulation through the cascaded modulators. This broadband multicarrier generator primarily comprises a $LiNbO_3$-based phase modulator, a Mach-Zehnder amplitude modulator, an microwave RF oscillator, and a tunable RF phase shifter.

When continuous wave light is fed into a phase-EOM, the phase-modulated light at the modulator output generates multiple sidebands at frequencies both below and above the optical frequency of the incident light, resulting in the generation of optical frequency comb (OFC). The frequency interval between adjacent comb lines is equal to the modulating RF frequency, and the number of sidebands is determined by the RF power and the half-wavelength voltage of the modulator, referred to as $V_\pi$. Typically, the optical power of the fundamental tone and the higher-order sideband harmonics can be described by the Bessel function, which means that the amplitude of the generated sidebands cannot remain uniform across the entire OFC spectrum. However, it is important to note that the light spectrum can be readily flattened by time-gating the phase-modulated light, thereby achieving hybrid phase and amplitude modulation. This process allows for a more uniform distribution of sideband amplitudes across the spectrum. The resulting electrical field from the hybrid modulation can be explicitly expressed as:

$$E_{out}(t) = e^{i2\pi v_c t} \cdot e^{iM_{dep}\sin(2\pi RFt)} \cdot \{\sin(2\pi RFt + \Delta\phi) + DC\} \quad (1)$$

where the first term represents the electrical field of the incident light, the second term corresponds to the phase modulation, and the third term accounts for the time gating of the phase-modulated light with the transmission position of the intensity modulator adjusted by the DC bias. The phase offset $\Delta\phi$ in the third term is the differential phase between the RF signals applied to the phase and amplitude modulators. The values of the phase offset and the DC bias position are crucial for the spectral tailoring of the OFC spectrum, as will be discussed in detail later. These parameters play a key role in achieving the desired spectral characteristics, allowing for precise control over the distribution and uniformity of the comb lines across the optical frequency comb spectrum.

## III. EXPERIMENTAL RESULTS AND SIMULATIONS ON THz WAVE GENERATION

In our experiment, a continuous wave (CW) laser operating at a wavelength of 1560 nm undergoes strong phase modulation at RF frequency of 14.5 GHz with an estimated modulation depth of ~9.6, followed by intensity modulation under conditions of $\Delta\phi=0$ and DC=1 (See (1)). Notice that the condition of DC=1 corresponds to the quadrature position of the intensity modulator transmission. The applied DC bias is highly stabilized using a feedback loop and the bias position in the transmission of the amplitude modulator can be set to any desired position. To obtain such large optical phase modulation, three identical phase-EOMs were cascaded in series. Two modulators were modulated at 33 dBm while the third was modulated at 30 dBm; thereby, the applied voltage was 76.5 $V_{pp}$ in total. According to the datasheet, the half-wavelength voltage, referred to as $V_\pi$, is about 4 V at 15 GHz. The RF phase synchronization between the four modulators is precisely controlled using external RF phase shifters placed in the corresponding RF arm of the modulator. This hybrid modulation technique generates a presumably flat optical frequency comb over a spectral bandwidth of approximately 7 nm, with the measured spectrum showing good agreement with the simulated spectrum obtained by (1), as shown in Fig.2(a). The key configuration for obtaining a flattened comb spectrum is to precisely select the time window where the laser experiences a negative frequency chirp, as shown in Fig.2(b). However, it is observed that the optical power at the edges of the OFC is higher than in the middle. This discrepancy is attributed to the sinusoidal phase modulation pattern, which differs from the ideal parabolic pattern that is inherently required to achieve a perfectly flat power distribution across the comb spectrum. But such nonlinearity can be mitigated by optimizing the operational parameters of the intensity modulator [19]. Nevertheless, the primary advantage of this configuration lies in its exceptional ability to generate ultra-short optical pulses. The induced negative frequency chirp across the optical pulse can be effectively mitigated by utilizing a dispersive optical medium that exhibits an opposite dispersion. This technique enables the initially chirped pulse to be compressed in the time domain, potentially reaching the transform-limited pulse width when the frequency chirp generated by the hybrid modulation remains perfectly linear [23].

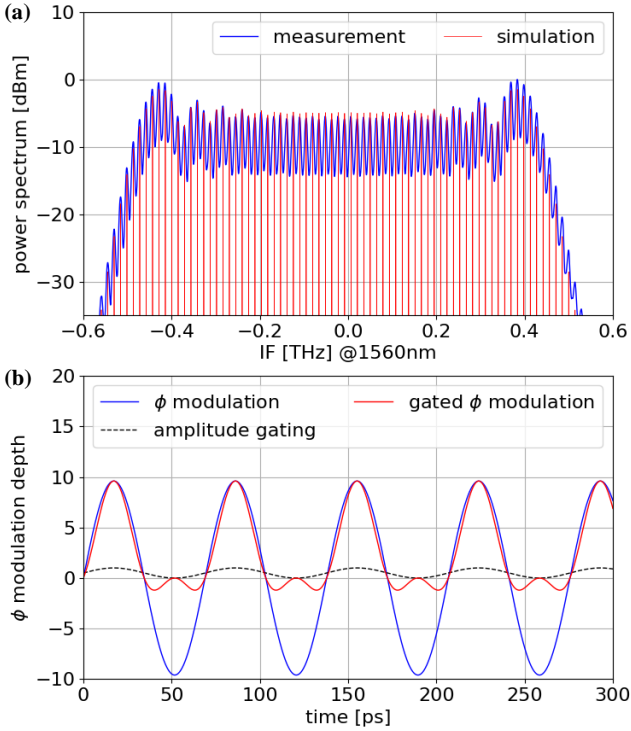

**Fig. 2** (a) Comparison between the measured and simulated OFC generated by the hybrid modulation, showing a good agreement. (b) Illustration of time gating of phase modulation, selecting only the time window where the laser experiences negative frequency chirp.

The temporal profile of the temporally compressed optical pulse was simulated numerically in relation to the dispersion parameter, as shown in Fig.3(a). This simulation helps determine the exact amount of dispersion needed to achieve the maximum temporal compression of the chirped pulse, essentially determined by the modulation depth at the phase-EOM. The result from the simulation explicitly indicates that the pulse width gradually decreases as the dispersion of the optical medium increases, reaching a minimum at a dispersion value of -3.58 ps/nm. However, when the dispersion exceeds this value, the pulse width begins to broaden, as expected.

Based on these findings, we chose a chirped fiber Bragg grating as the dispersive medium, which has a dispersion parameter of about -3.6 ps/nm, to generate the coherent optical frequency comb. The temporal profile of the compressed pulse was then accurately obtained using an autocorrelator, as shown in Fig.3(b). The pulse width was measured to be 930 fs, assuming a Gaussian shape, which closely matches the results of the simulated compressed pulse.

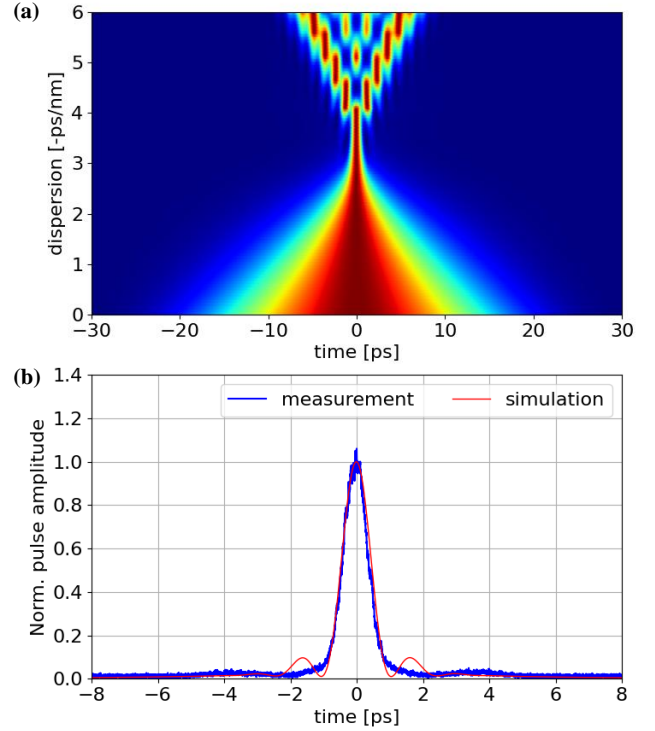

**Fig. 3** (a) Simulation of compressed optical pulse with respect to dispersion parameter. (b) Comparison between the measured and simulated compressed pulse with duration of ~930 fs, showing a good agreement.

However, it must be noticed that small adjustments to the operating parameters of the intensity modulator can potentially lead to the generation of an efficient THz wave pulse. This is because the output spectrum produced by the hybrid modulation is determined by the combination of two clock pulse spectra, which are influenced by the DC bias position used during amplitude modulation [24]. Fig.4 illustrates simulations results. When the intensity modulator is operated with $\Delta\phi=\pi/2$ and DC=0, as shown in Fig.4(a), the flattened comb spectrum can be transformed into a two color-like laser source (See Fig.4(b)) thanks to the significant optical power reduction in the middle part of the comb spectrum. Notice that the condition of DC=0 corresponds to the minimum position of the intensity modulator transmission. As a result, two distinct packages of comb lines are observed at the two extreme edges of the comb spectrum. However, due to the inevitable optical frequency chirp, the intensity profile of the output signal follows the intensity modulation pattern, as shown in Fig.4(c). According to our simulation results, all the spectral components within the remaining comb can become in phase by applying appropriated dispersion compensation, i.e., -5 ps/nm, resulting in the generation of a 3 ps optical pulse train that carries a terahertz wave at 0.754 THz.

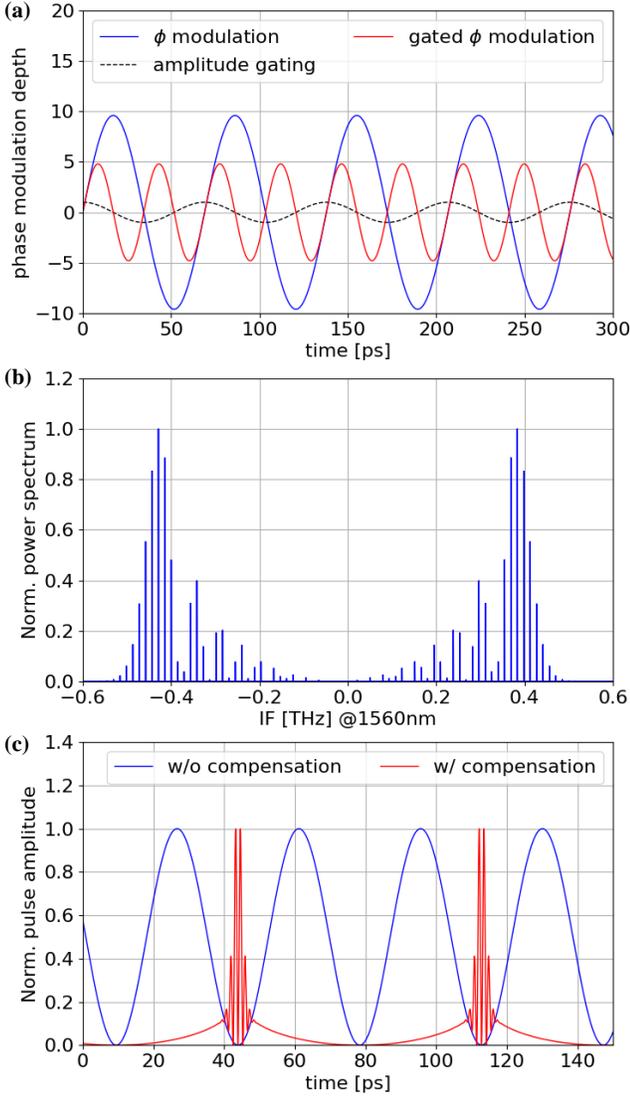

**Fig. 4** Simulation results. (a) Time gating of phase modulation, generating two clock pulses by the specific DC position of amplitude-EOM. (b) Tailored spectrum reducing the optical power of sidebands in the middle of the OFC spectrum. (c) Generation of 3 ps THz wave optical pulse train after applying a dispersion compensation of -5 ps/nm.

To validate the feasibility of the proposed technique, the operational parameters for generating electro-optic comb were first optimized to achieve the broadest comb spectrum. This step was necessary to ensure the accuracy and correctness of these operational parameters prior to the spectral tailoring on the comb spectrum. Therefore, the modulators were operated at the modulation frequency of 34.5 GHz, with the DC=1 and $\Delta\phi=0$. Note that in this experiment, only two phase modulators were utilized while 3 phase modulators were implemented in the previous experiment. RF powers of 33 dBm and 30 dBm were applied to the first and second modulators, respectively; thereby, the total applied voltage was 48.3 $V_{pp}$, generating an optical frequency comb over >THz spectral window. Fig.5(a) shows experimentally obtained frequency comb spectrum with numerically simulated one for a comparison purpose. Our simulation reveals that a modulation depth of 4.7 yields a good agreement between the two spectra. In addition, this theoretical value of modulation depth matches well the practical value applied to the modulator since the $V_\pi$ of the modulator is about 4.8 V at 34.5 GHz from the datasheet. In turn, the operation parameters to the amplitude-EOM were altered to $\Delta\phi=\pi/2$ and DC=0 to realize spectral tailoring, successfully demonstrating experimental result of generating two-tone-like laser source, as shown in the red profile in Fig.5(b). Looking closely at this spectrum, we can find that the fundamental comb line at the seed laser frequency is eliminated due to the unique feature of the DC bias operation condition.

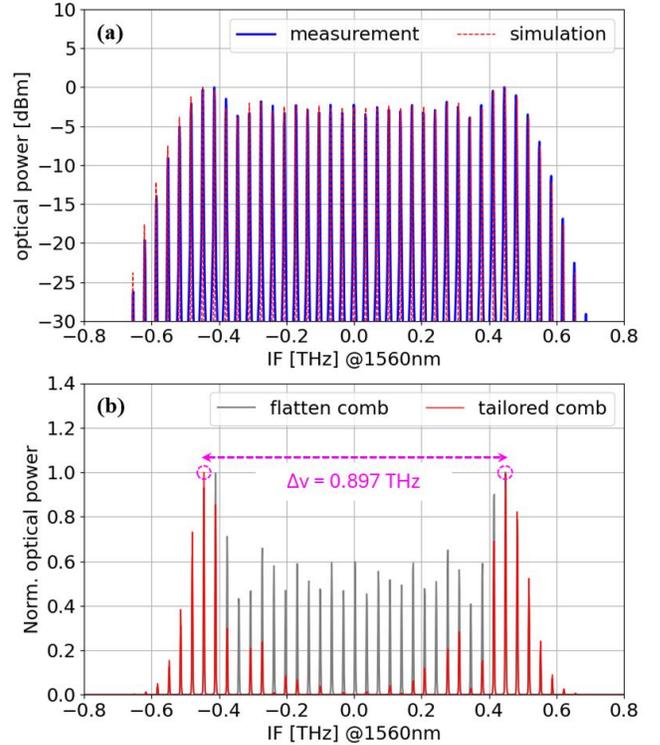

**Fig. 5** (a) Comparison between experimentally obtained OFC spectrum at modulation frequency of 34.5 GHz and simulation. (b) Comparison between experimentally obtained flat comb spectrum and tailored spectrum for efficient generation of THz wave at 0.897 THz by effectively suppressing optical powers in the middle part of the generated comb spectrum.

The primary purpose of increasing the modulation RF frequency in this experiment is to unambiguously evaluate the capability of maximum achievable carrier frequency of THz wave, in order words, the frequency tuning range of THz wave generated by this technique. As a matter of fact, this tunability is inherently constrained by the trade-off relation between the modulation frequency and modulation depth applied for the light phase modulation. In general, the $V_\pi$ of the electro-optic modulator increases with RF frequency due to several factors such as dielectric loss in the electro-optic modulator, the RF power attenuation in the electrodes and the mismatching between the velocities of the incident light in the waveguides and the modulating RF signal along the electrodes [25-27]. However, the outcomes from the two different operational conditions indicate that the frequency tuning range of THz wave generation remains very similar, resulting in measured frequency turnabilities of 0.754 THz and 0.897 THz at the modulation frequency of 14.5 GHz and 34.5 GHz, respectively. Due to a substantial difference in the modulation frequency, the peak sideband, which corresponds to the highest optical power across the spectrally tailored comb, was observed at the 26$^{th}$ harmonic and 13$^{th}$ harmonics for the two cases. This observation aligns well with theoretical expectations based on the difference in modulation depth and the fact that the increase in $V_\pi$ with modulation depth between 14.5 GHz and 34.5 GHz is relatively marginal. Overall, according to these experimental results we would like to emphasize that the modulation RF frequency has a significant impact on the maximal achievable carrier frequency of THz wave, generated by in this approach. Additionally, higher modulation frequencies provide larger spectral spacing between adjacent comb lines, which improves the effectiveness of optical filtering. This will be particularly beneficial when selecting specific sidebands, i.e., only two sidebands at the +13$^{th}$ and -13$^{th}$ harmonic comb lines; hence, enhancing the purity of the generated THz wave.

## V. CONCLUSIONS

In summary, a novel approach for generating tunable terahertz (THz) waves has been proposed and experimentally demonstrated. This technique utilizes hybrid modulation of light through cascaded phase- and amplitude-EOMs, offering an efficient method for THz wave generation. To validate the concept, a dual-frequency light source with a spectral spacing of 0.897 THz was successfully demonstrated. However, it is important to note that the performance of the generated two-tone laser is not fundamentally limited by physical constraints. Instead, the limitations stem from the current experimental setup, including the available bulk electro-optic components and the precision of the electrical control signals. This implies that with improved components, such as photonic integrated circuit-based electro-optic modulators having a lower half-wavelength voltage $V_\pi$, and more refined control mechanisms, the system could potentially achieve even better performance, surpassing the current experimental limits. In addition, this technique could be further optimized by employing more advanced modulation functions. For instance, using a triangular shape for phase modulation and a rectangular shape for amplitude modulation could lead to higher-quality terahertz signals, enhancing signal integrity and precision. These improvements would provide greater control over the characteristics of the generated THz waves, such as their spectral purity, frequency tunability, and phase noise.

We believe that the proposed method holds significant promise in simplifying the photonic architecture needed for THz wave generation. By reducing the complexity of the system while still achieving high-quality THz signals, this approach could pave the way for more practical and efficient THz wave generation techniques. In the future, with further development and refinement based on various PIC platform, this method could enhance the tunability, spectral purity, and stability of THz signals, making it a valuable tool for a wide range of applications, from scientific research to industrial uses such as imaging, communication, and spectroscopy.